\documentclass[a4paper]{article}

\usepackage{INTERSPEECH2021}
\usepackage{bm}
\usepackage{hyperref}
\usepackage{multirow}
\usepackage{multicol}
\usepackage{subcaption}
\usepackage{lipsum}
\usepackage{diagbox}
\usepackage{float}
\newcommand{\etal}{\emph{et al.}\ }

\title{Utilizing Self-supervised Representations for MOS Prediction}
\name{Wei-Cheng Tseng$^*$, Chien-yu Huang$^*$, Wei-Tsung Kao, Yist Y. Lin, Hung-yi Lee \thanks{$^*$ These authors contributed equally.}}
\address{
  College of Electrical Engineering and Computer Science, National Taiwan University, Taiwan}
\email{\{r09942094, r08921062, r09942067, r08922048, hungyilee\}@ntu.edu.tw}

\begin{document}

\maketitle
\begin{abstract}
Speech quality assessment has been a critical issue in speech processing for decades.
Existing automatic evaluations usually require clean references or parallel ground truth data, which is infeasible when the amount of data soars.
Subjective tests, on the other hand, do not need any additional clean or parallel data and correlates better to human perception.
However, such a test is expensive and time-consuming because crowd work is necessary.
It thus becomes highly desired to develop an automatic evaluation approach that correlates well with human perception while not requiring ground truth data.
In this paper, we use self-supervised pre-trained models for MOS prediction.
We show their representations can distinguish between clean and noisy audios.
Then, we fine-tune these pre-trained models followed by simple linear layers in an end-to-end manner.
The experiment results showed that our framework outperforms the two previous state-of-the-art models by a significant improvement on Voice Conversion Challenge 2018 and achieves comparable or superior performance on Voice Conversion Challenge 2016.
We also conducted an ablation study to further investigate how each module benefits the task.
The experiment results are implemented and reproducible with publicly available toolkits\footnote{\url{https://github.com/s3prl/s3prl}}.
\end{abstract}
\noindent\textbf{Index Terms}: MOS prediction, speech quality assessment, self-supervised learning

\section{Introduction}

Speech quality assessment is to evaluate the quality of audios, and it has been an important part of speech processing to measure the performance of a system for decades.
Several assessment metrics were used to evaluate different aspects of audio quality.
In speech enhancement, perceptual evaluation of speech quality \cite{rix2001perceptual} (PESQ) and short-time objective intelligibility \cite{taal2010short} (STOI) are widely used to measure the noise reduction.
In speech syntheses such as text-to-speech and voice conversion, mel cepstral distance \cite{kubichek1993mel} (MCD) is used to measure the distortion of synthesized speech.
These metrics require reference audio, which implies the need for clean, parallel data.

When reference audios are not available, the most common way to evaluate speech quality is the mean opinion score (MOS).
Each subject is asked to give the audios opinion scores, integers ranging from 1 to 5, and the MOS is the mean score of several subjects.
A higher score indicates better quality and vice versa. MOS correlates better to human perception compared to automatic assessments above.
However, such measurement usually requires a great number of humans to involve, making it time-consuming.
Several machine-learning-based models \cite{yoshimura2016hierarchical, fu2018quality, patton2016automos, avila2019non, 1709883} were thus proposed for automatic speech quality assessment.
Lo \etal \cite{lo2019mosnet} verified the predictability of MOS with statistical method \cite{efron1994introduction} and proposed MOSNet for MOS prediction.
MBNet \cite{leng2021mbnet} further utilizes the judge identities in the training dataset for modeling the bias of subjects.
The bias modeling improves the correlation between the predicted and real scores, enhancing the generalizability to unseen systems. However, these previous works only rely on scarce human-labeled data, which could limit the performance.

Self-supervised learning enables the model to learn meaningful representations from large-scale unlabeled data.
CPC \cite{oord2018representation} predicts the future representation in a contrastive learning manner.
Wav2vec2.0 \cite{baevski2020wav2vec} learns the representation by masking the latent space.
APC \cite{chung2019unsupervised} autoregressively generates the input feature in the next time step.
TERA \cite{liu2020tera} reconstructs the original complete input from a corrupted one.
Self-supervised learning has achieved remarkable performance in several tasks including automatic speech recognition \cite{baevski2020effectiveness}, speaker verification \cite{xia2020self}, voice conversion \cite{lin2020fragmentvc}, and so on.
However, their potential for speech quality assessment has not been explored yet.
This paper presents the first use of self-supervised representations pre-trained on large-scale unlabeled data for automatic MOS prediction.
We first showed that self-supervised pre-trained models can cluster audios in various types, and then proposed a new framework in which the model is fine-tuned with some simple yet effective modules.
Our framework surpasses the two previous state-of-the-art models on Voice Conversion Challenge (VCC) 2018 \cite{lorenzovoice} while being at least comparable on VCC 2016 \cite{toda2016voice}.
We also conducted an ablation study to see how each module affects performance.

\section{Preliminary Analysis}
\label{sec:preliminary}
Intuitively, if the representations from a self-supervised pre-trained model are more discriminative between the high- and low-quality audios than conventional features are, they are more likely beneficial to MOS prediction.
We thus first explore to what extent these models can evaluate audio quality without fine-tuning. 
Two datasets are involved in this paper: VCC 2016 and VCC 2018.
In the challenge, participants submit the audios generated by their voice conversion systems.
Then, to evaluate these systems, subjects were asked to score the synthesized audios from different systems based on the audio quality.
We refer to the mean scores from all subjects of an utterance as \textbf{utterance-level} score, and the average of utterance-level scores of a system is called \textbf{system-level} score.
In VCC 2016 dataset, only the system-level MOS is available.
VCC 2018 dataset, in contrast, provides detailed scoring information, where the score of each subject for each utterance is available.
Following previous work \cite{leng2021mbnet}, We divided VCC 2018 into training, validation, and testing set with 13580, 3000, and 4000 utterances respectively.
For VCC 2016 dataset, since only the system-level MOS score is available, we only used it in the testing stage.

We sampled synthetic data from the best and the worst systems as well as ground truth in VCC 2018.
Additionally, we also collected real-world noises from WHAM! \cite{wichern2019wham} and music from MUSAN \cite{snyder2015musan}.
Then we extracted their representations with wav2vec 2.0 base \cite{baevski2020wav2vec} and projected them to 2D space using t-SNE \cite{van2008visualizing}.
\begin{figure}
    \centering
    \includegraphics[width=.85\linewidth]{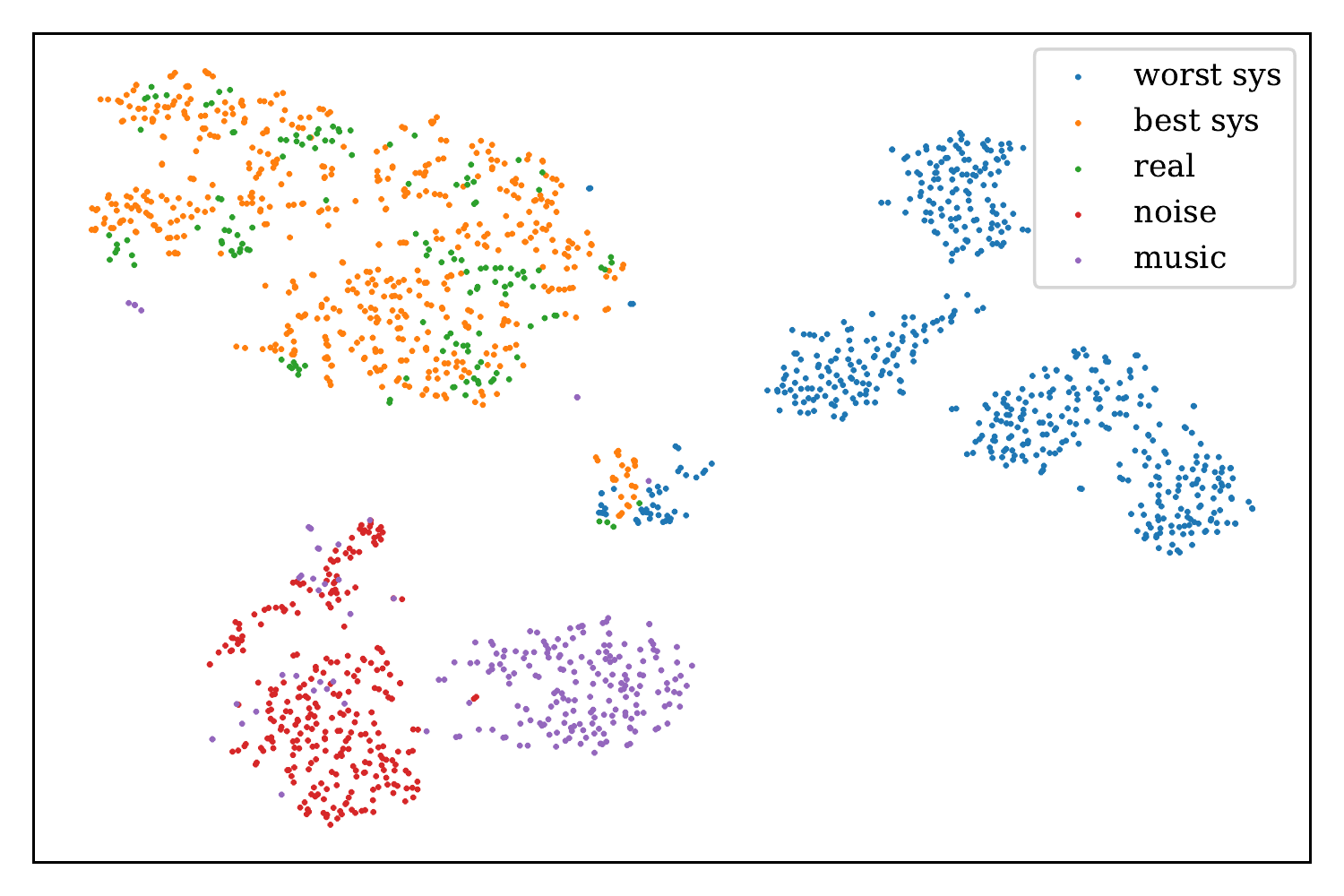}
    \caption{A visualization of representations of different kinds of audios extracted by wav2vec 2.0.}
    \label{fig:wav2vec2-scatter}
\end{figure}
Figure~\ref{fig:wav2vec2-scatter} illustrates the relationships among different kinds of audios.
We can see that real clean speech (green) clusters together, and audios from the best system (orange) are mixed with them.
On the other hand, audios from the worst system do not cluster with clean speech (blue far away from green), neither do noise (red) and music (purple).
This suggests wav2vec 2.0 can classify different kinds of audios.
Synthetic speech sounding more realistic tends to be closer to the cluster of real speech, and vice versa.

We extended the analysis by canonical correlation analysis \cite{thompson2005canonical} (CCA) on four pre-trained models: wav2vec 2.0 base, TERA base, APC with 3-layer GRU, and CPC.
CCA finds a linear transform mapping the representations to scores, and the linear correlation between these estimated scores and ground truth is maximized.
For each utterance, the model gives frame-level representations, which then form the utterance-level one by taking the average over all frames.
These utterance-level representations were then used for CCA.
We performed CCA on VCC 2018 training set and then applied the found transform on VCC 2018 testing set and VCC 2016 to see whether the representations are general.
As the baseline, we included two conventional speech features in the analysis: MFCC and Mel-spectrogram.

Table~\ref{tab:cca-result} presents the linear correlation between estimated scores and ground truths on VCC 2018 testing set and VCC 2016 using four different self-supervised pre-trained models.
We can see that the correlation on these representations is much higher than the baseline, which implies they contain information about audio quality.
Also, when the found transform is applied on VCC 2016, the correlation remains much high, suggesting that these representations are general to evaluate audio quality.
With both qualitative and quantitative analysis, we can know that it is useful and feasible to use self-supervised pre-trained models for MOS prediction.

\begin{table}[!h]
    \centering
    \caption{
    The linear correlation coefficients between estimated scores and ground truths.
    Estimated scores were obtained using the linear transform found with CCA on VCC 2018 training set.
    }
    \begin{tabular}{l|c|c|c}
         \hline
         \multirow{2}{*}{Model} & Utterance-level & \multicolumn{2}{c}{System-level} \\
         \cline{2-4}
         {} & VCC 2018 & VCC 2016 & VCC 2018 \\
         \hline
         \hline
         wav2vec 2.0 & 0.734 & 0.966 & 0.99 \\
         TERA & 0.727 & 0.943 & 0.987 \\
         APC & 0.678 & 0.891 & 0.964 \\
         CPC & 0.699 & 0.890 & 0.980 \\
         \hline
         MFCC & 0.183 & 0.196 & 0.326 \\
         Mel-spec. & 0.215 & 0.487 & 0.618 \\
         \hline
    \end{tabular}
    \label{tab:cca-result}
\end{table}

\section{Proposed Framework}
\begin{figure}[H]
    \centering
    \includegraphics[width=.9\linewidth]{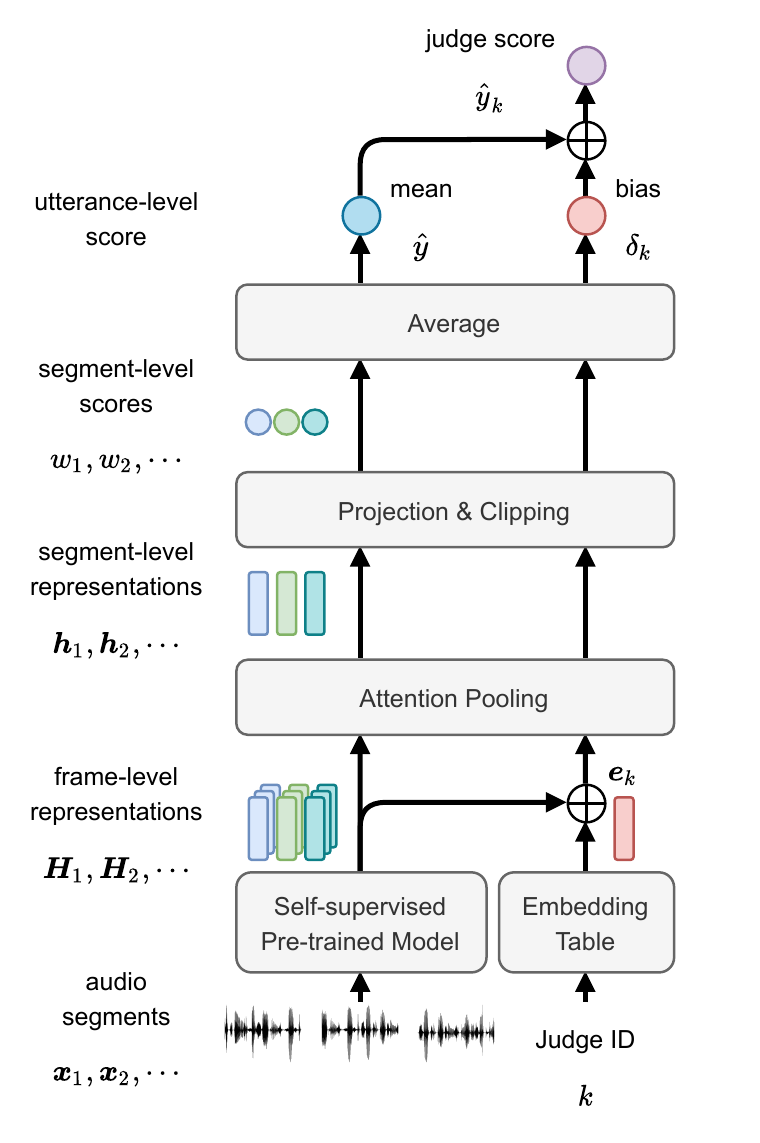}
    \caption{
    The proposed framework for MOS prediction.
    }
    \label{fig:proposed_framework}
\end{figure}

In MOS test, an audio $\bm{x}$ is evaluated by $K$ subjects giving a sequence of scores $y_1, y_2, \cdots, y_{K}$.
These scores are averaged to obtain a mean score, denoted as $y$.
Our goal is to predict $y$ from $\bm{x}$ with a pre-trained model.
Figure~\ref{fig:proposed_framework} explains our framework, which consists of segmental embeddings \cite{1521436, rybach2009audio, chung2016audio, wang2018segmental} of pre-trained model (Sec.~\ref{sec:pre-trained-models}), attention pooling (Sec.~\ref{sec:attention-pooling}), bias network (Sec.~\ref{sec:bias-network}), and range clipping (Sec.~\ref{sec:range-clipping}).

\subsection{Segmental embeddings}
\label{sec:pre-trained-models}
We start with a pre-trained self-supervised model denoted as $f(\cdot)$.
An audio of $T$ sample points $\bm{x} = [x_1, x_2, \cdots, x_T]$ is first divided into several segments $\bm{x}_{seg}$,
\begin{align}
    &\bm{x}_{seg} = [\bm{x}_1, \bm{x}_2, \cdots, \bm{x}_N],\\
    &\bm{x}_{i} = [x_{i \cdot S}, x_{i \cdot S + 1}, \cdots, x_{i \cdot S + \ell}],
\end{align}
where $\ell$ is the length of each segment and $S$ is the stride.
Then the pre-trained model encodes these segments into a sequence of representations,
\begin{align}
    &f(\bm{x}_{seg}) = [\bm{H}_{1}, \bm{H}_{2}, \cdots, \bm{H}_{N}], \\
    &\bm{H}_i = [\bm{h}_{i1}, \bm{h}_{i2}, \cdots, \bm{h}_{iM}],
\end{align}
where $N$ is the number of segments and $M$ is the length of features of a segment.
Here $\bm{h}_{ij} \in \mathbb{R}^{d}$ is referred to as \textbf{frame-level} representations, and $d$ is their dimensionality.

\subsection{Attention pooling}
\label{sec:attention-pooling}
Attention mechanism \cite{vaswani2017attention} has gained much success in several tasks.
Here we use attention pooling \cite{safari2020self} to obtain segment-level representation from a sequence of frame-level representations.
For each segment, the attention module first encodes frame-level representations into queries,
\begin{align}
    \bm{Q}_{i} = \text{softmax}(\bm{W}\bm{H}_{i}),
    \label{eq:attention-query}
\end{align}
where $\bm{H}_i \in \mathbb{R}^{d \times M}$, $\bm{W} \in \mathbb{R}^{1 \times d}$, $\bm{Q}_{i} \in \mathbb{R}^{1 \times M}$.
It then gives the \textbf{segment-level} representation by
\begin{align}
    \bm{h}_{i} = \bm{H}_{i}\bm{Q}_{i}^T.
    \label{eq:attention-pooling}
\end{align}
We then obtain the segment-level score $w$ by $g(\cdot): \mathbb{R}^d \rightarrow \mathbb{R}$,
\begin{align}
    w_{i} = g(\bm{h}_{i}).
    \label{eq:segment-score}
\end{align}
Finally, the utterance-level score $\hat{y}$ is determined by taking the average of segment-level scores,
\begin{align}
    \hat{y} = \frac{1}{N}\sum^{N}_{i = 1}w_{i}
    \label{eq:uttr-score}
\end{align}

\subsection{Bias network}
\label{sec:bias-network}
Utilizing individual judge's scores in training has been shown to improve the performance \cite{leng2021mbnet}.
Here we also adopt a bias network to make use of them.
A judge ID $k$ is first transformed into an embedding $\bm{e}_{k}$ by a trainable embedding table.
It is added to frame-level representations $\bm{h}_{ij}$ to form judge-biased representations $\hat{\bm{h}}_{ij}$.
Then, we obtain the bias of a judge $\delta_{k}$ from $\hat{\bm{h}}_{ij}$ by a bias network with similar process described in (\ref{eq:attention-query}) - (\ref{eq:uttr-score}).
We can determine the score given by judge $k$ by
\begin{align}
    \hat{y}_{k} = \hat{y} + \delta_{k}.
\end{align}
In the test stage, we use only the mean score and discarded the bias network as was done in MBNet for better performance.

\subsection{Range clipping}
\label{sec:range-clipping}
While the model is trained to fit the human scoring distributions, the existence of outliers is inevitable.
To reduce such distortion, we can apply hyperbolic tangent function to ensure a fixed range of segment-level score.
In this way, (\ref{eq:segment-score}) above becomes:
\begin{align}
    \hat{w}_i = 2\tanh{g(\bm{h}_i)} + 3.
    \label{eq:clip-segment-score}
\end{align}
The range of $\hat{w}_{i}$ is constrained between 1 and 5, and therefore the model is always guaranteed to give reasonable scores.
This clipping is only applied in the mean score prediction but not the bias ones because the bias is not necessarily a positive number.

\subsection{Training objectives}
\label{sec:training-objectives}
During the training, we minimize utterance- and segment-level mean squared error (MSE) for mean score.
As for the biased score for each judge, we minimize the MSE on utterance-level only.
The training objective is thus expressed as
\begin{align}
    \mathcal{L} = (\hat{y} - y)^2 + \frac{\alpha}{N}\sum^{N}_{i = 1}(w_{i} - y)^2 + \frac{\beta}{K}\sum^{K}_{j = 1}(\hat{y}_{k} - y_{k})^2,
    \label{eq:uttr-seg-loss}
\end{align}
where $\alpha$ and $\beta$ are hyperparameters balancing the losses.

Although Lo \etal\cite{lo2019mosnet} showed frame-level MSE loss mitigates the variance of predicted scores within an utterance, here we resort to the segment-level MSE instead.
We believe that it is more reasonable to score segments of utterance but not every frame, as the former one correlates better to human perception.

\begin{table*}[!h]
    \caption{
    The performances of our frameworks (trained with/without bias network) and baselines on VCC 2016 and VCC 2018 test set.
    }
    \centering
    \begin{subtable}[!h]{.39\textwidth}
    \centering
    \caption{utterance-level}
    \begin{tabular}{l|c|c|c}
        \hline
         \multirow{2}{*}{Model} & \multicolumn{3}{c}{VCC 2018} \\
         \cline{2-4}
         {} & MSE & LCC & SRCC \\
         \hline
         \hline
         \multicolumn{4}{c}{\textbf{with} bias network} \\
         \hline
         wav2vec 2.0 & 0.450 & \textbf{0.739} & \textbf{0.718} \\
         TERA & 0.496 & 0.705 & 0.680 \\
         APC & 0.425 & 0.685 & 0.659 \\
         CPC & \textbf{0.408} & 0.698 & 0.668 \\
         \hline
         MBNet & 1.134 & 0.628 & 0.592 \\
         \hline\hline
         \multicolumn{4}{c}{\textbf{without} bias network} \\
         \hline
         wav2vec 2.0 & 0.496 & 0.722 & 0.698 \\
         TERA & 0.426 & 0.692 & 0.661 \\
         APC & 0.440 & 0.678 & 0.650 \\
         CPC & 0.423 & 0.686 & 0.651 \\
         \hline
         MOSNet & 0.471 & 0.639 & 0.604 \\
         \hline
    \end{tabular}
    \label{tab:uttr-level-results}
    \end{subtable}
    \hfill
    \begin{subtable}[!h]{.59\textwidth}
    \centering
    \caption{system-level}
    \begin{tabular}{l|c|c|c|c|c|c}
        \hline
        \multirow{2}{*}{Model} & \multicolumn{3}{c|}{VCC 2016} & \multicolumn{3}{c}{VCC 2018} \\
        \cline{2-7}
        {} & MSE & LCC & SRCC & MSE & LCC & SRCC \\
        \hline
        \hline
        \multicolumn{7}{c}{\textbf{with} bias network} \\
        \hline
        wav2vec 2.0 & 0.483 & \textbf{0.964} & 0.893 & 0.083 & 0.981 & 0.968 \\
        TERA & 0.655 & 0.940 & \textbf{0.916} & 0.090 & 0.987 & 0.972 \\
        APC & 0.318 & 0.941 & 0.886 & 0.028 & 0.967 & 0.961 \\
        CPC & 0.180 & 0.957 & 0.854 & \textbf{0.016} & 0.981 & 0.968 \\
        \hline
        MBNet & \textbf{0.106} & 0.945 & 0.878 & 0.771 & 0.982 & 0.980 \\
        \hline\hline
        \multicolumn{7}{c}{\textbf{without} bias network} \\
        \hline
        wav2vec 2.0 & 0.615 & 0.961 & 0.868 & 0.104 & \textbf{0.991} & \textbf{0.981} \\
        TERA & 0.407 & 0.942 & 0.896 & 0.023 & 0.988 & 0.977 \\
        APC & 0.238 & 0.938 & 0.887 & 0.046 & 0.971 & 0.949 \\
        CPC & 0.170 & 0.955 & 0.844 & 0.016 & 0.983 & 0.960 \\
        \hline
        MOSNet & 0.336 & 0.901 & 0.850 & 0.054 & 0.960 & 0.918 \\
        \hline
    \end{tabular}
    \label{tab:system-level-results}
    \end{subtable}
    \label{tab:comparison_all}
\end{table*}

\begin{table*}[h!]
    \caption{
    The system-level performance on VCC 2016 in \textbf{with} scenario when one of the modules is removed.
    The notation "- X" means module X is removed from the framework.
    }
    \centering
    \begin{tabular}{l|c|c|c|c|c|c|c|c|c|c|c|c}
        \hline
        \multirow{2}{*}{Model} & \multicolumn{3}{c|}{original} & \multicolumn{3}{c|}{- segmental embeddings} & \multicolumn{3}{c|}{- attention pooling} & \multicolumn{3}{c}{- range clipping} \\
        \cline{2-13}
        {} & MSE & LCC & SRCC & MSE & LCC & SRCC & MSE & LCC & SRCC & MSE & LCC & SRCC \\
        \hline
        \hline
        wav2vec 2.0 & 0.483 & 0.964 & 0.893 & 0.414 & 0.958 & 0.854 & 0.499 & 0.962 & 0.859 & 0.623 & 0.967 & 0.860 \\
        TERA & 0.655 & 0.940 & 0.916 & 0.660 & 0.887 & 0.785 & 0.637 & 0.941 & 0.920 & 0.509 & 0.934 & 0.905 \\
        APC & 0.318 & 0.941 & 0.886 & 0.373 & 0.944 & 0.902 & 0.306 & 0.938 & 0.890 & 0.364 & 0.926 & 0.866 \\
        CPC & 0.180 & 0.957 & 0.854 & 0.208 & 0.940 & 0.824 & 0.188 & 0.958 & 0.866 & 0.243 & 0.948 & 0.848 \\
        \hline
    \end{tabular}
    \label{tab:ablation}
\end{table*}

\section{Experimental Settings}
Following Section~\ref{sec:preliminary}, we used four self-supervised pre-trained models in the experiments: wav2vec 2.0, CPC, TERA, and APC.
These models were pre-trained with large-scale unlabeled data such as LibriSpeech \cite{7178964} and Libri-Light \cite{librilight} and can be accessed via publicly-available toolkits S3PRL \cite{S3PRL}.

For segmental embeddings, the duration ($\ell$) and stride ($S$) are 1.0 and 0.5 seconds respectively.
The raw representations $f(\bm{x}_{seg})$ from the pre-trained model were projected to 256-dim space for each frame.
Then, the projection $g(\cdot)$ from segment-level representations to scores was simply a linear layer.
For the bias network, we adopted similar architecture.

Two baseline models were included in the experiments for comparison: MOSNet and MBNet.
For MOSNet, we trained the model with the official implementation\footnote{\url{https://github.com/lochenchou/MOSNet}}.
As for MBNet, we resorted to an unofficial implementation\footnote{\url{https://github.com/sky1456723/Pytorch-MBNet}} because the official one is not available.
These models were trained with hyperparameters from the original papers.

We fine-tuned pre-trained models for 20k steps in three learning rates: $1\mathrm{e}{-4}$, $5\mathrm{e}{-5}$ and $1\mathrm{e}-5$ with a warm-up in the first 500 steps and linear decay in the remaining steps.
For baseline models, we followed the settings in the original papers.
We performed testing on the validation set every 250 steps, and the checkpoint with the best system-level performance (in terms of Spearman's rank correlation coefficient) was then used for evaluation on the testing set.

\section{Results}
\subsection{Quantitative results}
We evaluated the performance on VCC 2016 and VCC 2018 test set in three metrics: mean squared error (MSE), linear correlation coefficient \cite{pearson1920notes} (LCC), and Spearman's rank correlation coefficient \cite{spearman1904proof} (SRCC).
MSE measures the absolute difference between predicted scores and ground truths, while the latter two tell how correlated the predictions and ground truths are.

Table~\ref{tab:comparison_all} lists the performances of the proposed framework using different pre-trained models along with two baselines.
We compared our framework with the baselines in two scenarios: \textbf{with} and \textbf{without} bias network.
In \textbf{with} scenario, our framework is trained with the help of a bias network and compared to MBNet, which also utilizes the individual judge scores.
In \textbf{without} scenario, we train the model without bias network ($\beta = 0$ in (\ref{eq:uttr-seg-loss})) and take MOSNet as the baseline, which simply uses the mean score of each utterance.
From Table~\ref{tab:uttr-level-results}, we see in utterance-level MOS, all pre-trained models outperformed the baseline models in both LCC and SRCC significantly.
In terms of MSE, only wav2vec 2.0 in \textbf{without} scenario slightly fell behind MOSNet.
We can also observe that the models in \textbf{with} scenario outperformed those in \textbf{without} scenario pairwisely, though the bias network was not used in the testing.

On the other hand, Table~\ref{tab:system-level-results} presents the system-level performance for different models on VCC 2016 and VCC 2018 testing set.
We see that the LCC and SRCC of all models including baselines are higher than those in utterance-level, which means system-level scores are more predictable than those in utterance-level.
In \textbf{without} scenario, all the pre-trained models surpassed MOSNet in terms of LCC and SRCC on VCC 2016 and 2018, while wav2vec 2.0 and TERA fell behind MOSNet in MSE.
However, we believe that LCC and SRCC better evaluate the performance than MSE does.
This is because MSE cannot capture the positive/negative correlation between two sets of samples but only reflects the absolute difference.
As in \textbf{with} scenario, MBNet achieved the best SRCC in VCC 2018 (even better than it was reported in the original paper though trained with less audios).
However, when it comes to SRCC in VCC 2016, pre-trained models performed better except CPC.
As for LCC, we see all pre-trained models are at least comparable to MBNet in both VCC 2016 and 2018, and wav2vec 2.0 even surpassed much (LCC = 0.964).
Last, we can again find that \textbf{with} scenario is superior to \textbf{without} scenario, suggesting that using individual judge scores is an important key to achieving better MOS prediction.

\subsection{Ablation study}
We then inspected how each module works by removing them one at a time.
In the case segmental embedding is not used, we directly calculate the utterance score from frame-level representations.
Also, following the previous work \cite{lo2019mosnet}, the segment-level MSE in (\ref{eq:uttr-seg-loss}) becomes frame-level MSE, which enforces all frames produce the same score.
When attention pooling is removed, a simple mean pooling is adopted for segment-level scores.
Last, without the presence of range clipping, the range of $w$ is no longer limited and therefore can be arbitrary real number.

Table~\ref{tab:ablation} lists system-level performances on VCC 2016 in \textbf{with} scenario when each of the module is removed from our framework.
Due to the space limit, we do not present the results in \textbf{without} scenario, where a similar trend was observed.
We see as segment embedding is not used, the performance dropped for almost all models, showing that segment is a better unit for quality assessment than the frame is, as we mentioned in Sec~\ref{sec:training-objectives}.
Then, when attention pooling is replaced, there is little change.
This is probably because we adopted segment-level MSE in the training objective, which enforces all segments in an utterance give similar or same scores and therefore weaken the power of attention.
Last, we can observe that without range clipping, MSE and LCC worsened for almost all models.
This means the clipping works well for stabilizing model output within a reasonable range.
Meanwhile, we also see that SRCC was not affected much because SRCC only considers the ranking order of two sets, not the exact values in them.

\section{Conclusions}
This paper presents the first use of self-supervised pre-trained models for MOS prediction.
We show that these models can measure audio quality without fine-tuning in both qualitative and quantitative aspects, and then propose a framework in which the model is fine-tuned with simple yet effective modules.
Our framework outperformed the previous works on VCC 2018 and achieved comparable or superior performance on VCC 2016.
Experimental results showed that wav2vec 2.0 achieved the best or competitive performance in both utterance- and system-level.
We also conducted a thorough ablation study to explore how each module in the framework works.
The use of segmental embeddings boosts the performance the most, confirming that it is more reasonable to score a segment of utterance instead of every frame in the previous work.

\section{Acknowledgements}
We acknowledge the support of AWS Machine Learning Research Awards program.

\bibliographystyle{IEEEtran}

\bibliography{main}


\end{document}